\documentclass[aps,prb,reprint,superscriptaddress,showpacs,amsmath,amssymb,floatfix,tighten]{revtex4-1}
\usepackage{graphicx}
\usepackage{bm}

\usepackage[table]{xcolor}
\usepackage{multirow}

\usepackage{mathptmx}
\usepackage[breaklinks=true,colorlinks=true,urlcolor=blue,
citecolor=blue,linkcolor=blue,bookmarks=false]{hyperref}

\newcommand{\ltsim}{\protect\raisebox{-0.5ex}{$\:\stackrel{\textstyle <}{\sim}\:$}}
\newcommand{\gtsim}{\protect\raisebox{-0.5ex}{$\:\stackrel{\textstyle >}{\sim}\:$}}

\begin{document}

\title{Electronic structure of cyanocobalamin: DFT+QMC study}

\author{Selma Mayda$^{1,2}$, Zafer Kandemir$^{1}$, and Nejat Bulut}


\affiliation{Department of Physics, Izmir Institute of Technology, 
Urla 35430, Turkey\\
$^2$ Department of Materials Science and Engineering, 
Izmir Institute of Technology, Urla 35430, Turkey}

\date{\today}

\begin{abstract}
We study the electronic structure and the magnetic correlations of cyanocobalamin 
(C$_{63}$H$_{88}$CoN$_{14}$O$_{14}$P) by using the framework of the multi-orbital single-impurity Haldane-Anderson model 
of a transition metal impurity in a semiconductor host. Here, we first determine the parameters of the Anderson  
Hamiltonian by performing density functional theory (DFT) calculations. Then,
we use the quantum Monte Carlo (QMC) technique to obtain the electronic structure and the magnetic 
correlation functions for this effective model. We find that new electronic states, which 
correspond to impurity bound states, form above the lowest unoccupied level of the host semiconductor. These new states derive from 
the atomic orbitals at the cobalt site and the rest of the molecule. We observe that magnetic 
moments develop at the Co($3d_{\nu}$) orbitals and over the surrounding sites.
We also observe that antiferromagnetic correlations exist between the Co$(3d_{\nu})$ orbitals and 
the surrounding atoms. These antiferromagnetic correlations 
depend on the filling of the impurity bound states.
\end{abstract}


\maketitle

\section{Introduction}
\label{intro}
Vitamin B$_{12}$ is a very important organometallic molecule for biological systems \cite{Prada,Harris,Stich}. 
In this paper, we study the electronic structure of cyanocobalamin (CNCbl) which is a form of vitamin B$_{12}$. In Fig. 1,
we illustrate the molecular structure of CNCbl. 
The cobalt atom neighbours five nitrogen atoms, of which four are located in the corrin ring. 
The CN ligand is also attached to Co, making the rare cobalt-carbon bonding. The charge neutral CNCbl molecule has 718 electrons.
For this molecule, 
the HOMO (highest occupied molecular orbital) and the LUMO (lowest unoccupied molecular orbital) levels 
are separated by an energy gap of $\approx 2.2$ eV \cite{Firth}. The photoabsorption spectrum of CNCbl exhibits 
distinct peaks at $\approx 3.5$ eV and $4.5$ eV, of which origin remains elusive \cite{Firth}. In additon, 
it is known that CNCbl has a weak diamagnetic response \cite{Grun,Diehl}. 
\begin{figure}[!htp]
\includegraphics[width=8cm]{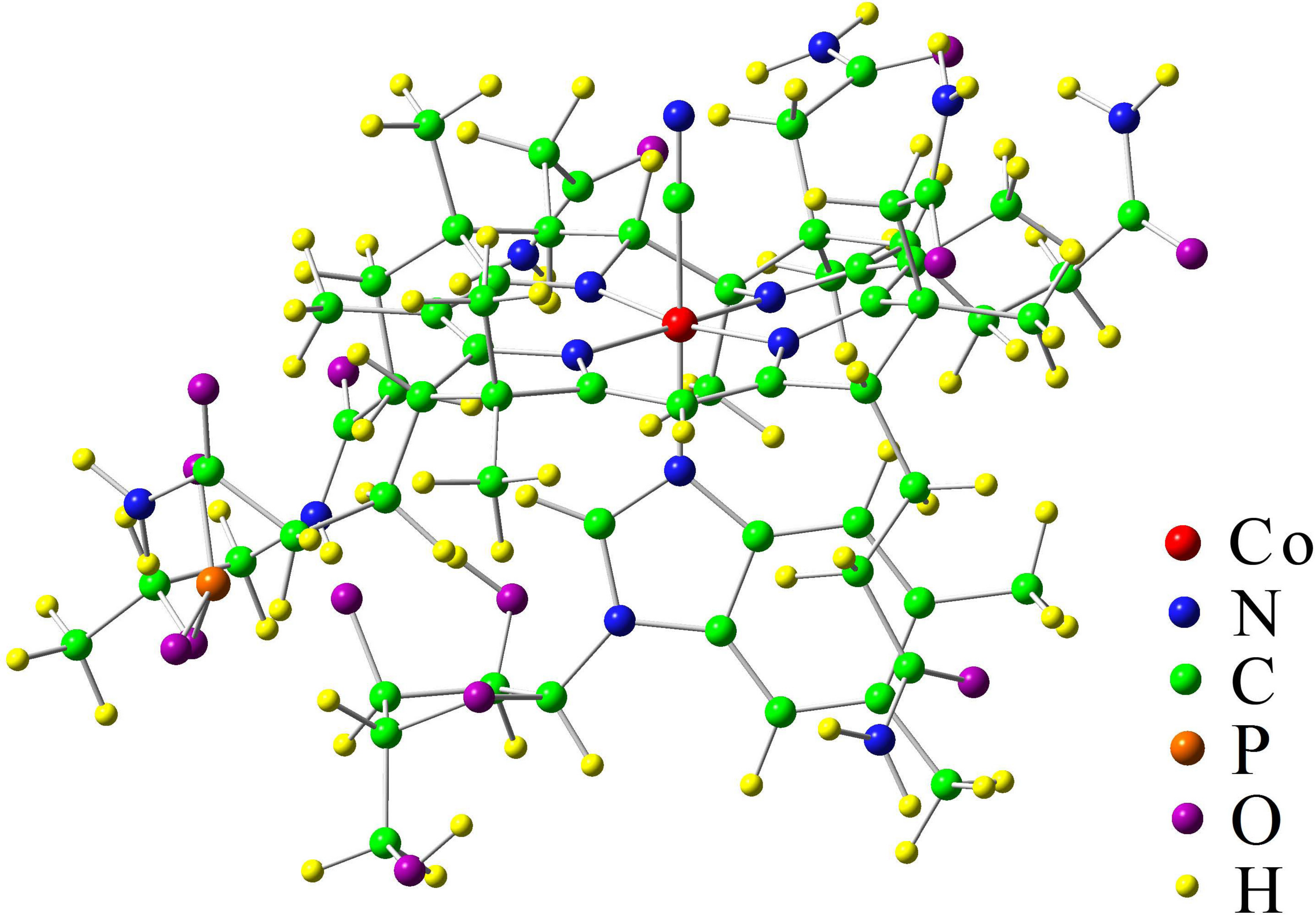}
\caption{(Color online) 
Schematic plot of the molecular structure of cyanocobalamin
(C$_{63}$H$_{88}$CoN$_{14}$O$_{14}$P), which contains 181 atoms.
} 
\label{fig1}
\end{figure}

Despite many years of research, there remains questions about the electronic structure and 
the functioning of vitamin B$_{12}$ as well as the role of the transition-metal cobalt atom. 
In this paper, we study the electronic structure of CNCbl from the perspective of many body physics. In 
particular, we use the combined density functional theory (DFT) and quantum Monte Carlo (QMC) approach to study 
the electronic structure and magnetic correlations of this molecule within the framework of the Haldane-Anderson 
model \cite{Haldane}. This model was initially introduced to describe the electronic state of Au in a semiconductor 
Ge host. Here, we use the Haldane-Anderson model because CNCbl exhibits a semiconductor energy gap and contains the 
transition-metal cobalt atom. 

In the combined DFT+QMC approach, we first determine the parameters of the Anderson Hamiltonian \cite{Anderson} 
by the DFT calculations carried out with the Gaussian program \cite{Gaussian}. Then, we study this effective Anderson Hamiltonian 
by performing QMC simulations with the Hirsch-Fye \cite{Hirsch} algorithm. In our calculations, we use an orbital independent intra-orbital Coulomb interaction $U$ 
 at the Co($3d_{\nu}$) orbitals and we neglect the inter-orbital Coulomb interactions along with the Hund's coupling. 
We take the value of $U$ to be 4 eV. In addition, the Co($3d_{\nu}$) energy 
levels are shifted in order to prevent the double counting of the local Coulomb interactions 
by both DFT and QMC. 

The outline of this paper is as follows. In Section 2, we obtain the one-electron parameters of the effective 
Anderson Hamiltonian. We present the QMC data on the electronic structure 
and the magnetic correlation functions in Section 3. Here, we find that impurity bound states develop above the LUMO level for CNCbl
instead of being inside the semiconducting energy gap. This is due to the discrete single-particle spectrum 
of the molecule. 
We observe that magnetic moments develop at the Co($3d_{\nu}$) orbitals and at the host states 
which have the strongest hybridization. We also observe that there are antiferromagnetic 
correlations between these host magnetic moments and the Co($3d_{\nu}$) moments.
These antiferromagnetic correlations disappear when the impurity bound states are filled.
We note that in Ref. [\onlinecite{Kandemir2}], we combined the Hartree-Fock (HF) approximation with the QMC technique 
to study CNCbl. In the HF+QMC calculations, we used only the smaller imidazole part  of 
the CNCbl instead of the 
whole molecule.  
In Section 4, we compare the HF+QMC results on Im-[Co$^{\rm III}$(corrin)]-CN$^+$ with the DFT+QMC results on CNCbl. 
Section 5 gives the summary and conclusions of the paper.

%
%
\section{Effective Anderson Hamiltonian}
\label{sec:1}
The multi-orbital single-impurity Anderson Hamiltonian \cite{Anderson} 
is given by 
\begin{eqnarray}
H &=& \sum_{m,\sigma} (\varepsilon_m-\mu) c^{\dagger}_{m\sigma} c_{m\sigma} +
\sum_{\nu,\sigma} (\varepsilon_{d\nu}-\mu) 
d^{\dagger}_{\nu\sigma} d_{\nu\sigma}  \nonumber \\
&+& \sum_{m,\nu,\sigma} ( V_{m\nu} c^{\dagger}_{m\sigma} d_{\nu\sigma} + 
V^*_{m\nu} d^{\dagger}_{\nu\sigma} c_{m\sigma} ) \\
&+& \sum_{\nu} U_{\nu} 
n_{\nu\uparrow} n_{\nu\downarrow} \nonumber
\label{hamiltonian}
\end{eqnarray}
where $c^{\dagger}_{m\sigma}$ ($c_{m\sigma}$) creates (annihilates) an electron 
in host state $m$ with spin $\sigma$, 
$d^{\dagger}_{\nu \sigma}$ ($d_{\nu \sigma}$) is the creation
(annihilation) operator for a localized electron with spin $\sigma$ at 
the Co($3d_{\nu}$) orbital, and 
$n_{\nu\sigma}= d^{\dagger}_{\nu\sigma} d_{\nu\sigma}$.
Here, $\varepsilon_m$ and $\varepsilon_{d\nu}$ are the 
energies of the host and the Co($3d_{\nu}$) impurity states, respectively,
The hybridization matrix element 
between these states is $V_{m\nu}$. The intra-orbital Coulomb repulsion is $U_{\nu}$.
Finally, a chemical potential $\mu$ is introduced  
since the QMC calculations are performed in the grand canonical ensemble.

We obtain the one-electron parameters $\varepsilon_{m}$, $\varepsilon_{d \nu}$ and 
$V_{m \nu}$ as explained below. 
Within DFT, the one-electron wave function $\psi_{n}({\bf r})$ are determined from 
 \begin{eqnarray}
  F({\bf r})\psi_{n}({\bf r})=E_n\psi_{n}({\bf r}),
  \label{dftt}
 \end{eqnarray}
where the Kohn-Sham operator \cite{Kohn} is 
 \begin{eqnarray}
\nonumber F({\bf r}) = \left(-\dfrac{\hbar^{2}}{2m_{e}}\nabla_{{\bf r}}^{2}+V_{ext}({\bf r}) + \int d^{3}r' \dfrac{\rho({\bf r'})}{\lvert {\bf r}-{\bf r'}\rvert}+ V_{xc}({\bf r}) \right)\\
\label{dft}
\end{eqnarray}
and the molecular orbital energy is $E_{n}$. 
The molecular orbitals $\psi_{n}$ can be expanded in terms of the $N$ atomic orbitals,
\begin{eqnarray}
| \psi_{n} \rangle = \sum_{i}^{N} C_{ni} | \phi_{i} \rangle,
\label{psi}
\end{eqnarray}
where $C_{ni}$ are the elements of the coefficient matrix $\bf{C}$. Substituting Eq. (4) in 
Eq (2), the Roothan equation, $\bf {CF=ECS}$, is obtained. Here, the elements of the Kohn-Sham 
matrix $\bf{F}$ are defined as 
\begin{eqnarray}
F_{ij} = \int d^{3}  r\, \phi^{*}_{i}({\bf r})\, F({\bf r})\, \phi_{j}({\bf r})
\end{eqnarray}
and the overlap matrix  $\bf{S}$ has the matrix elements $S_{ij} = \langle \phi_{i} | \phi_{j} \rangle$.
Because the atomic orbitals do not form an orthogonal basis, we use the 
natural atomic orbitals (NAO's) \cite{Reed} which form an orthogonal basis. The NAO's form a maximally localized basis set. 
Next, we express the Kohn-Sham matrix in the NAO basis. We take the Co($3d_{\nu}$) NAO's as the impurity orbitals and 
their energy levels as $\varepsilon_{d \nu}$'s in the Anderson Hamiltonian. Diagonalizing the remaining part of the 
Kohn-Sham matrix, we obtain the host eigenstates $|u_m\rangle$ and their energy levels $\varepsilon_{m}$ in additon 
to the hybridization matrix elements $V_{m \nu}$
This procedure is explained in more detail in Ref. [\onlinecite{Kandemir2}].
We use the Gaussian program \cite{Gaussian} with the BP86 energy functional \cite{BP86} and the 6-31G basis set with $N=1035$ basis functions  
to obtain the DFT solutions.

\begin{figure} 
{\includegraphics[width=6.8cm]{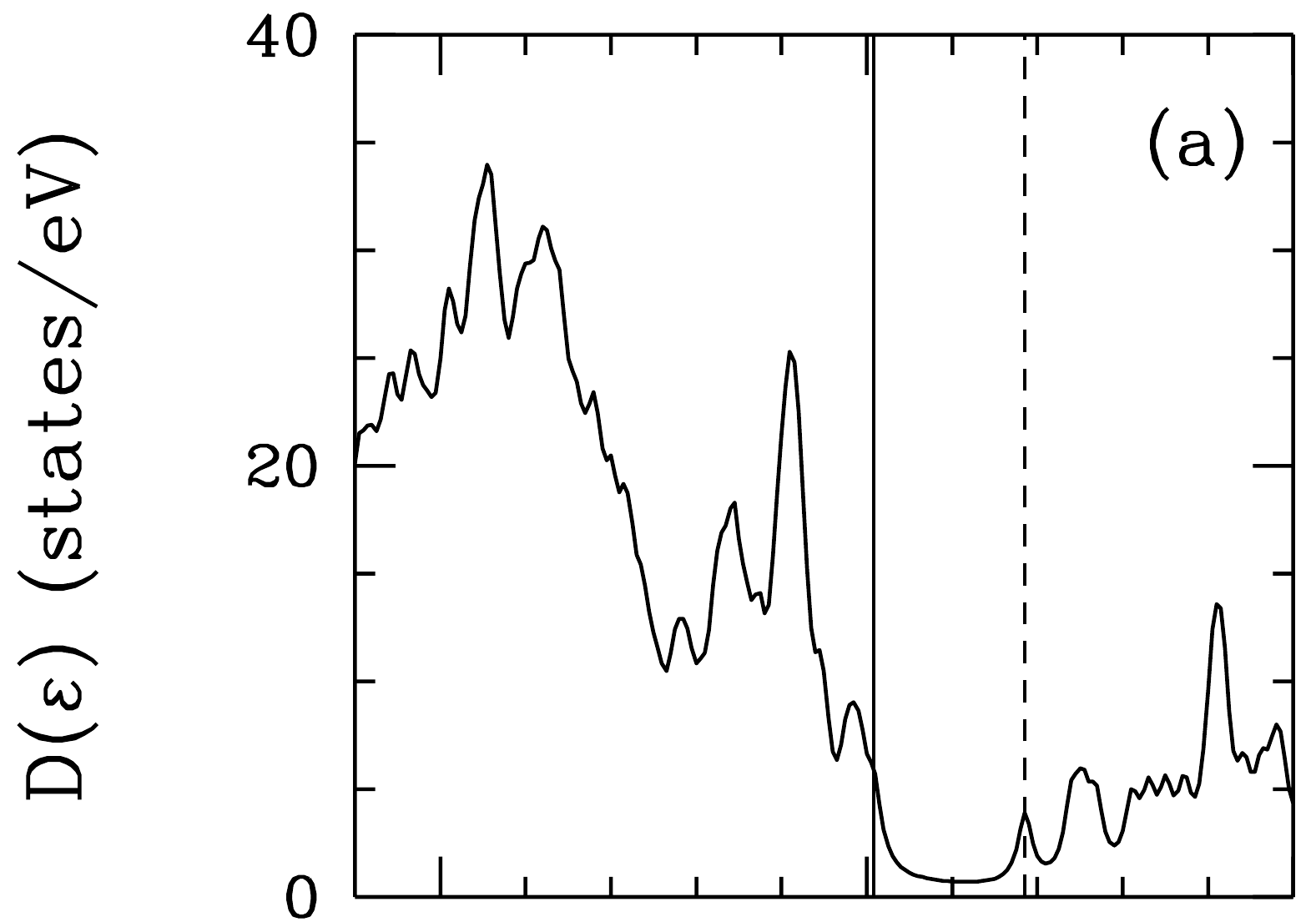}}
{\includegraphics[width=6.8cm]{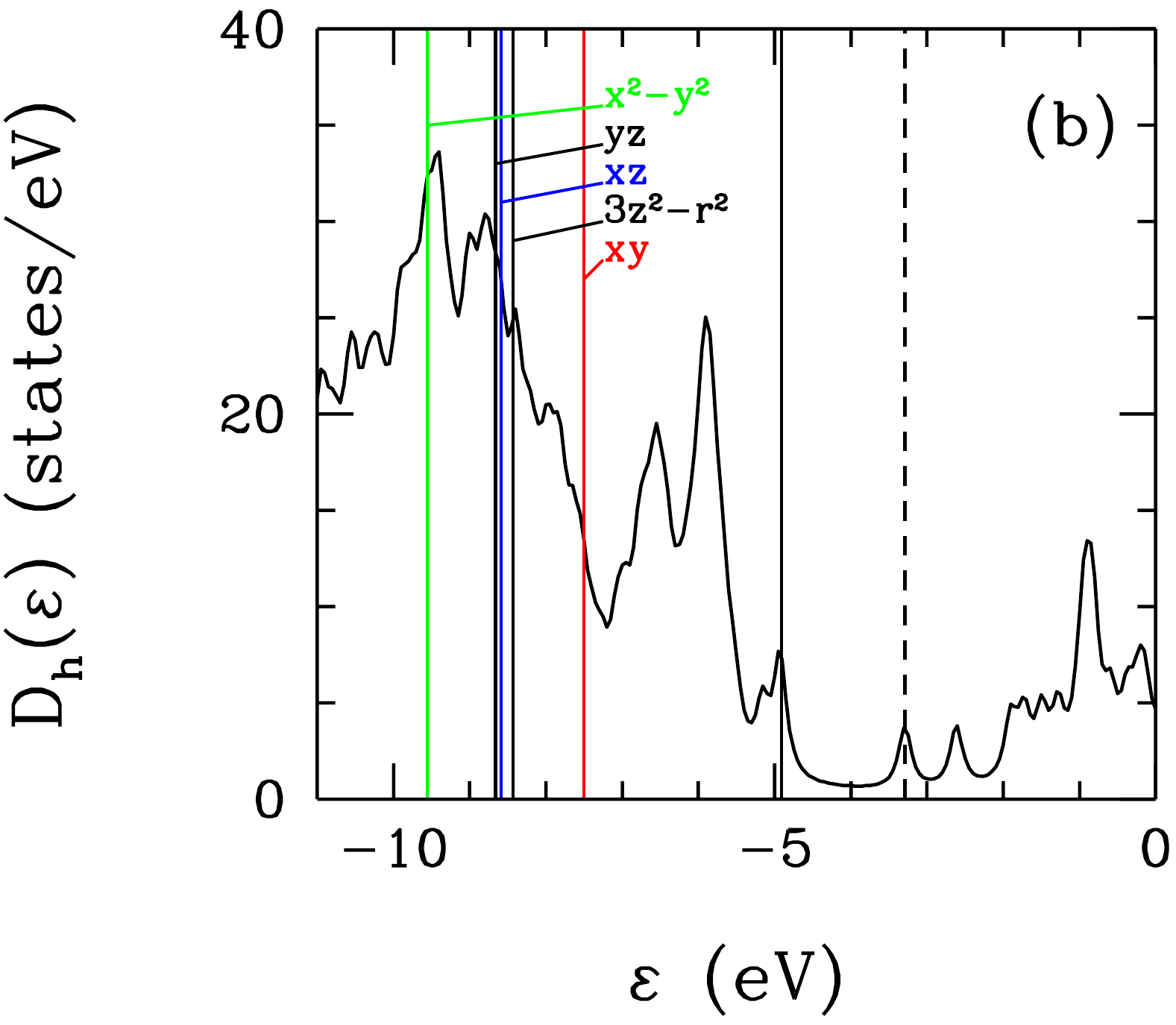}}
\caption{(Color online) 
(a) Density of states $ D(\varepsilon) $ of CNCbl 
obtained by the DFT calculations.
(b) Density of states of the host states of the effective Haldane-Anderson model
$D_h(\varepsilon)$. The shifted Co($3d_{\nu}$) natural atomic orbital levels $\tilde{\varepsilon}_{d\nu}$ are also indicated as vertical lines.
Here, the Co($3d_{\nu}$) levels have been shifted by $\mu^{\mathrm {DC}}_{\nu}$, which was obtained 
for $U=4$ eV. The vertical solid and dashed 
lines denote the HOMO and LUMO levels, respectively.
}
\label{fig2}
\end{figure}

In the QMC calculations, we use $U=4$ eV. In various Co compounds, the intra-orbital 
Coulomb interaction at the Co($3d_{\nu}$) orbitals is estimated to be between 4 eV and 
5 eV \cite{Sasioglu}. It is important to note that, in the DFT+QMC approach, 
the on-site Coulomb interaction $U$ is taken into 
account both by the DFT and the QMC calculations. Therefore, in order to prevent this double counting, an orbital-dependent 
double-counting term $\mu_{\nu}^{\rm {DC}}$, which is defined as 
\begin{eqnarray}
\mu^{\mathrm {DC}}_{\nu} = \frac{U \langle n^{\mathrm{DFT}}_{d\nu} \rangle}{2}
\end{eqnarray}
is substracted from the bare Co($3d_{\nu}$) levels,
$\varepsilon_{d\nu} 
\rightarrow \tilde{\varepsilon}_{d\nu} = \varepsilon_{d\nu}  - \mu_{\nu}^{\rm {DC}}$
\cite{Anisimov,Czyzyk,Kunes,Karolak}. In the Anderson Hamiltonian, $\tilde{\varepsilon}_{d\nu}$ is used instead of $\varepsilon_{d\nu}$. Here, $\langle n^{\mathrm{DFT}}_{d\nu} \rangle$ 
is the electron number in the Co($3d_{\nu}$) NAO's obtained by the DFT calculations. 
 
We begin presenting data by showing the density of states  $D(\varepsilon) = \sum_{n=1}^N \delta(\varepsilon-E_n)$ in Figure \ref{fig2}(a).
In this figure, the highest occupied molecular orbital (HOMO) is located at -4.9 eV,
and the lowest unoccupied molecular orbital (LUMO) is located at -3.2 eV, which means that 
the energy gap is 1.7 eV.
In Fig. \ref{fig2}(b), 
the host density of states $D_h(\varepsilon) = \sum_{m=1}^{N-5} \delta(\varepsilon-\varepsilon_m)$
is shown as a function of energy $\varepsilon$. In this figure, vertical lines indicate the 
shifted Co($3d_{\nu}$) NAO energy levels $\tilde{\varepsilon}_{d\nu}$. 
For these parameters,
while the $\nu=x^2-y^2 $ NAO is located at $ \varepsilon \approx -9.5$ eV, 
the $ \nu=yz $, $ xz $ and $ 3z^2-r^2 $ NAO's are located at $ \varepsilon \approx -8.5$ eV and 
the $ \nu=xy $ NAO is located at $ \varepsilon \approx -7.5$ eV.
We note that here, we choose a coordinate system in which the x and y axis are located at 45 degrees to the Co-N bond direction 
instead of being parallel.

\begin{figure}
\includegraphics[width=6.8cm]{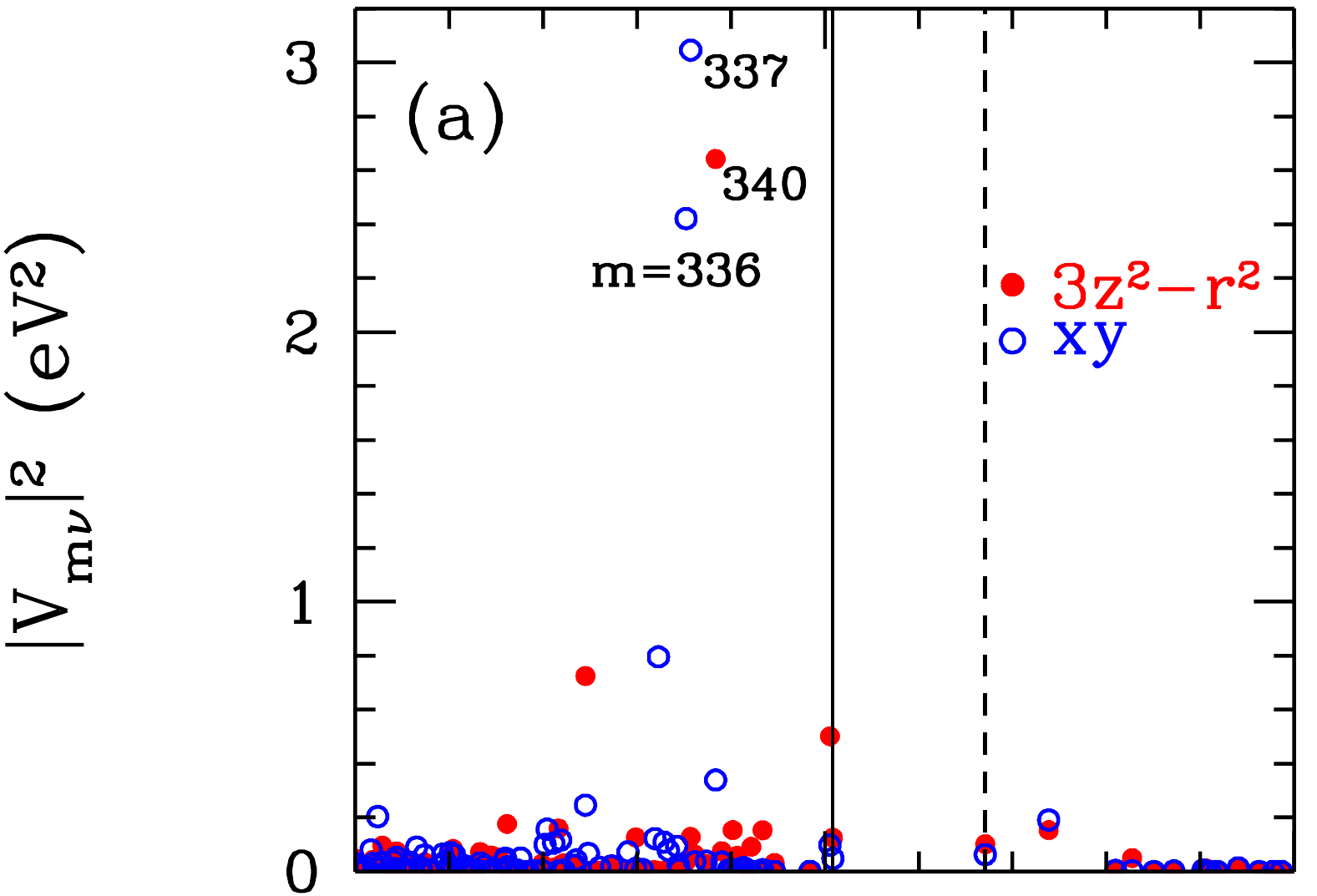} 
\includegraphics[width=6.8cm]{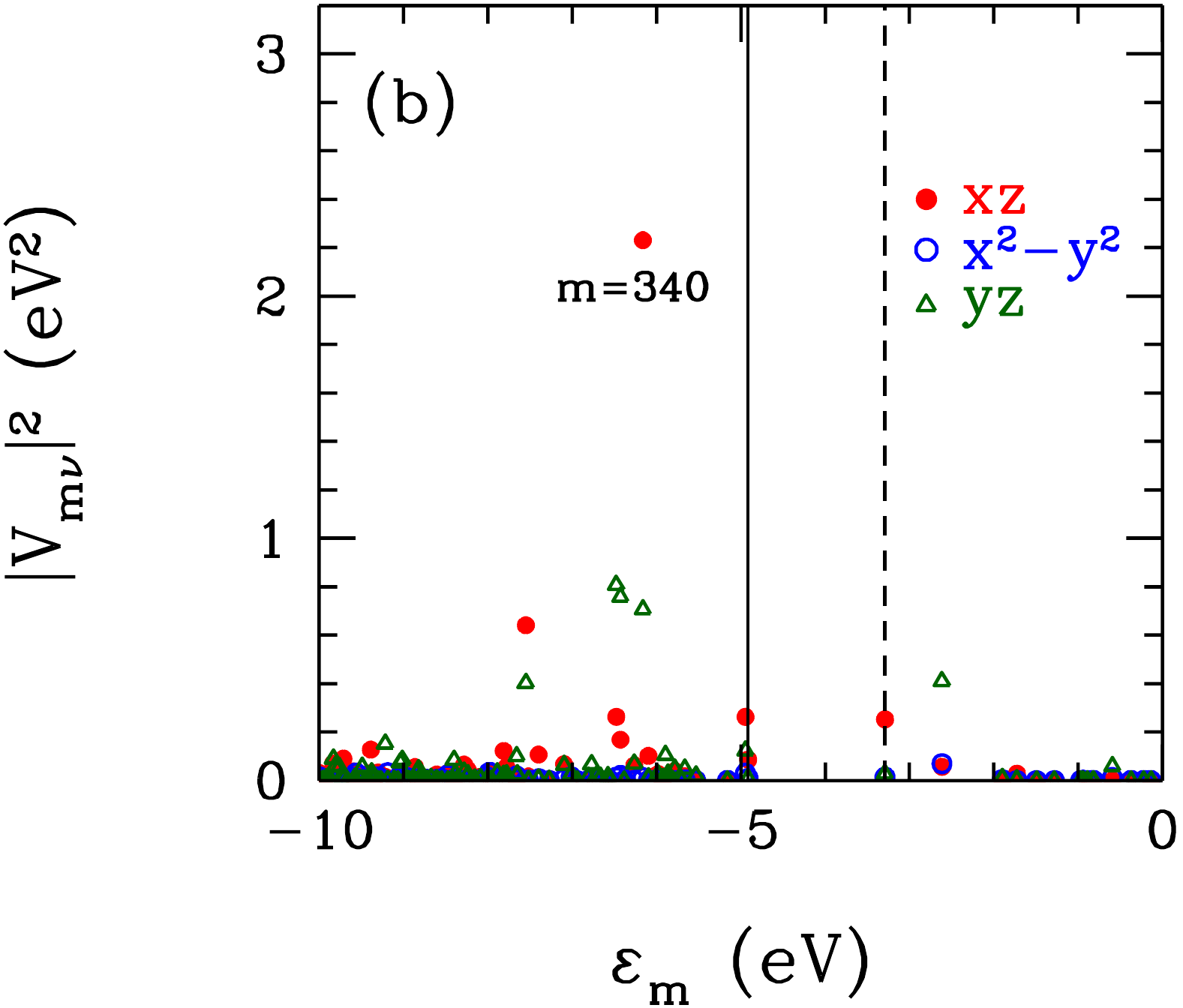}
\caption{(Color online) 
DFT results on the 
square of the hybridization matrix elements $ |V_{m\nu}|^2 $ 
between the Co($3d_{\nu}$) natural atomic orbitals and the $m$'th host states 
versus the host energy $ \varepsilon_{m} $. In (a) 
results are shown for the $3d_{3z^2-r^2} $ and $ 3d_{xy} $ natural atomic orbitals, and 
in (b) for the $3d_{xz} $, $ 3d_{x^2-y^2} $ and $ 3d_{yz} $ natural atomic orbitals. 
Here, the vertical solid and dashed lines denote the values of the 
HOMO and LUMO levels, respectively.
We observe that 
$m=336$, $337$ and $340$'th host states have the strongest 
hybridization matrix elements. 
}
\label{fig3}
\end{figure}

The DFT data on 
the square of the hybridization matrix elements $|V_{m\nu}|^2$ between the $m$'th host eigenstate $|u_m\rangle$ and the Co($3d_{\nu}$) NAO's 
are shown as a function of $\varepsilon_m$ in Fig. \ref{fig3}.
Here, we observe that the $m=336, 337$ and $340$ host states have the largest hybridization matrix 
elements. The $m=336$ and $337$ host states hybridize most strongly with the Co$(3d_{xy})$ NAO, while, 
the $m=340$ host state has the strongest hybridization with the Co$(3d_{3z^2-r^2})$ and Co$(3d_{xz})$ NAO's. 
In the QMC data, we will see that these host states are strongly influenced by the local Coulomb interaction.
 
\begin{figure}[t]
\centering
\includegraphics[width=5.8cm]{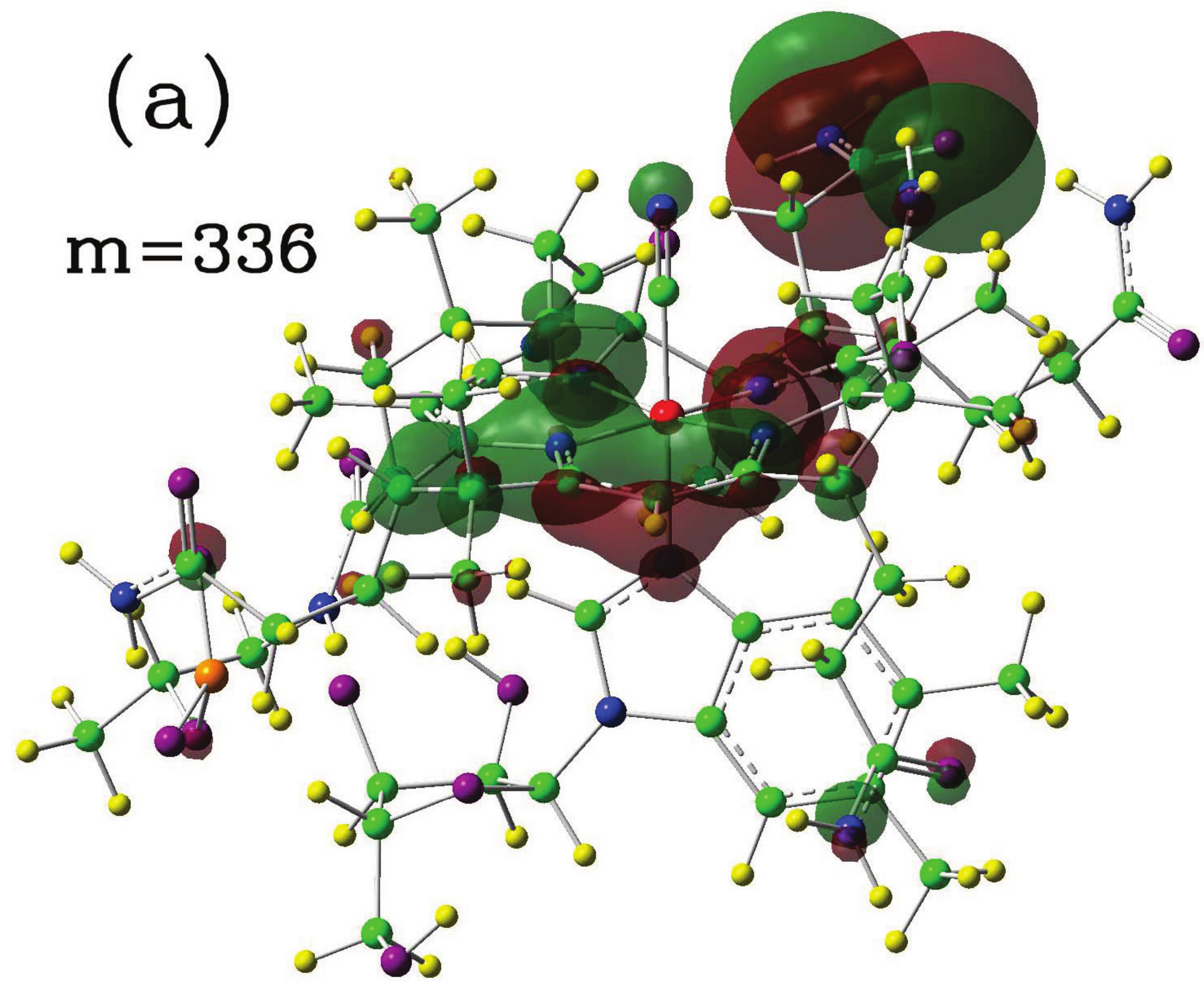} 
\includegraphics[width=5.8cm]{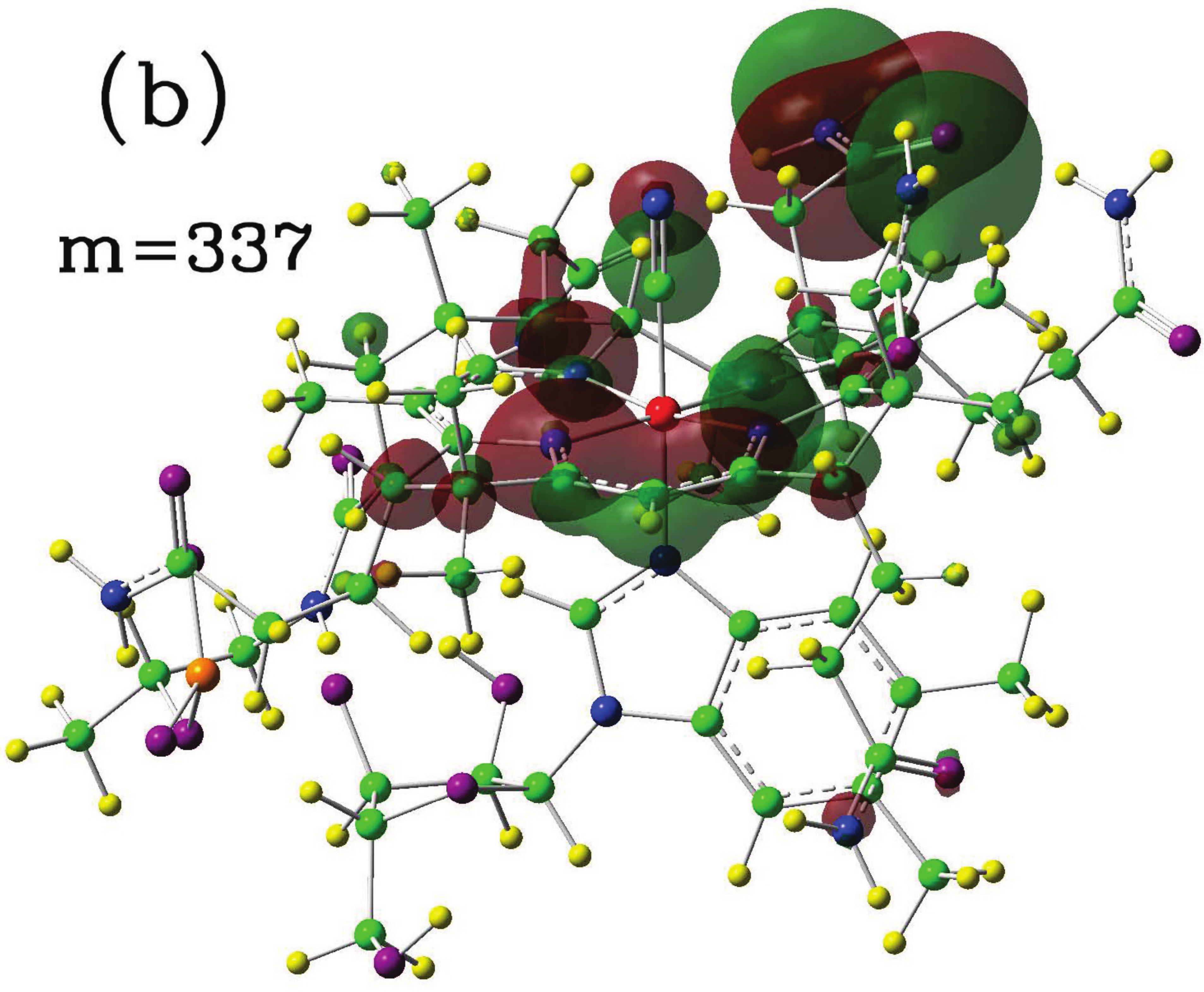} 
\includegraphics[width=5.8cm]{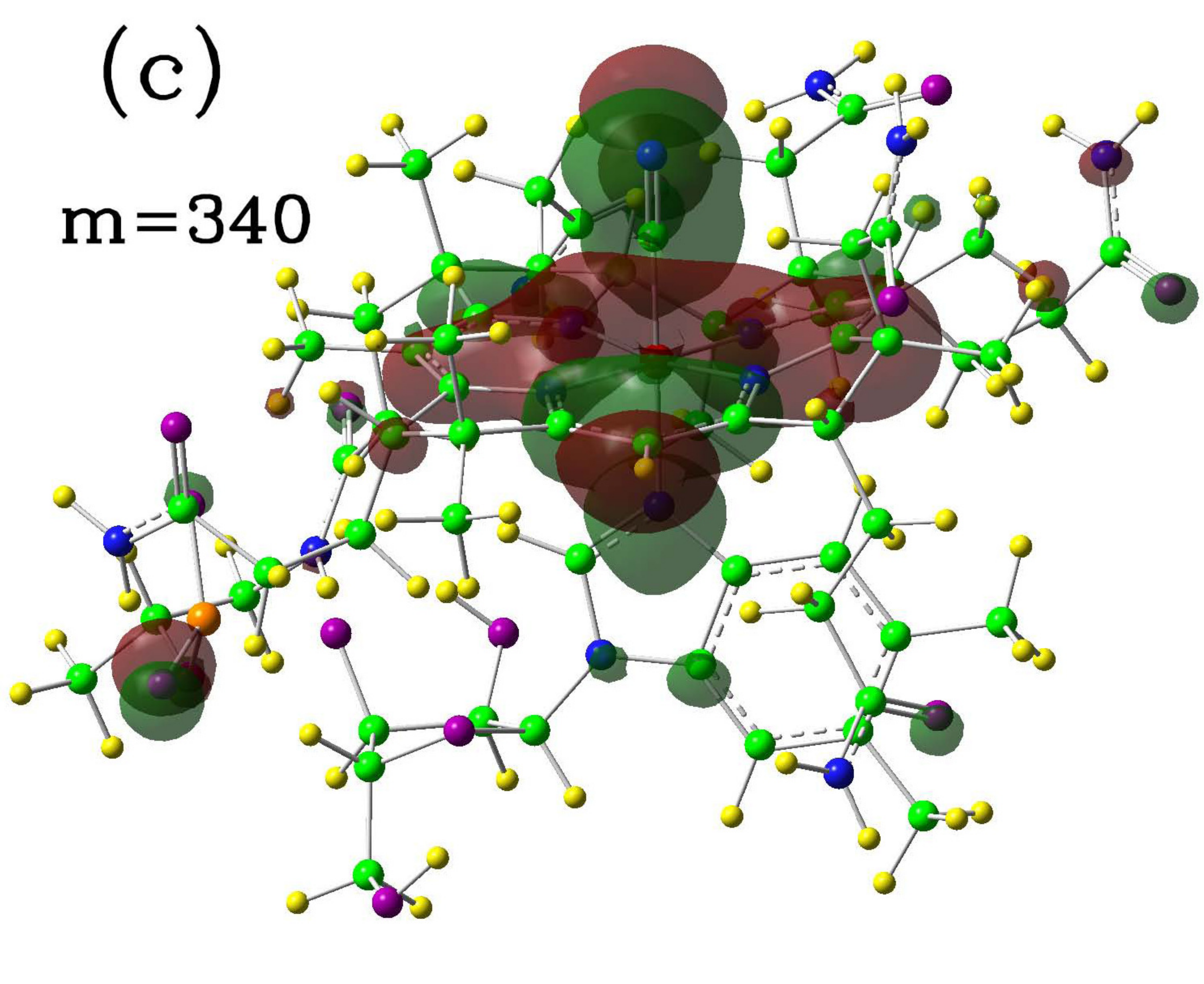} 
\caption{(Color online) 
Illustration of the $ m=336$, $337$ and $340$'th host states in terms of the natural atomic orbitals. 
}
\label{fig4}
\end{figure}

In order to gain insight into these host states, in Fig. \ref{fig4} we illustrate 
the $m=336, 337$ and $340$ host states in terms of the NAO's. These host states contain contributions from NAO's around the Co site. 
In particular, the $m=340$'th host state contains significant amount of weight from the CN ligand.

\section{Quantum Monte Carlo results}
\label{sec:2}
In this section, QMC data on the effective 
Haldane-Anderson model for CNCbl are presented. For this model, QMC calculations
were performed by using the Hirsch-Fye QMC algorithm \cite{Hirsch}. 
In the QMC calculations, a discrete Matsubara time step of 
$ \Delta\tau = 0.13$ eV$^{-1} $ is used. The results are presented for temperature $ T=700 $ K 
in the grand canonical ensemble. 

\begin{figure}
\includegraphics[width=6.8cm]{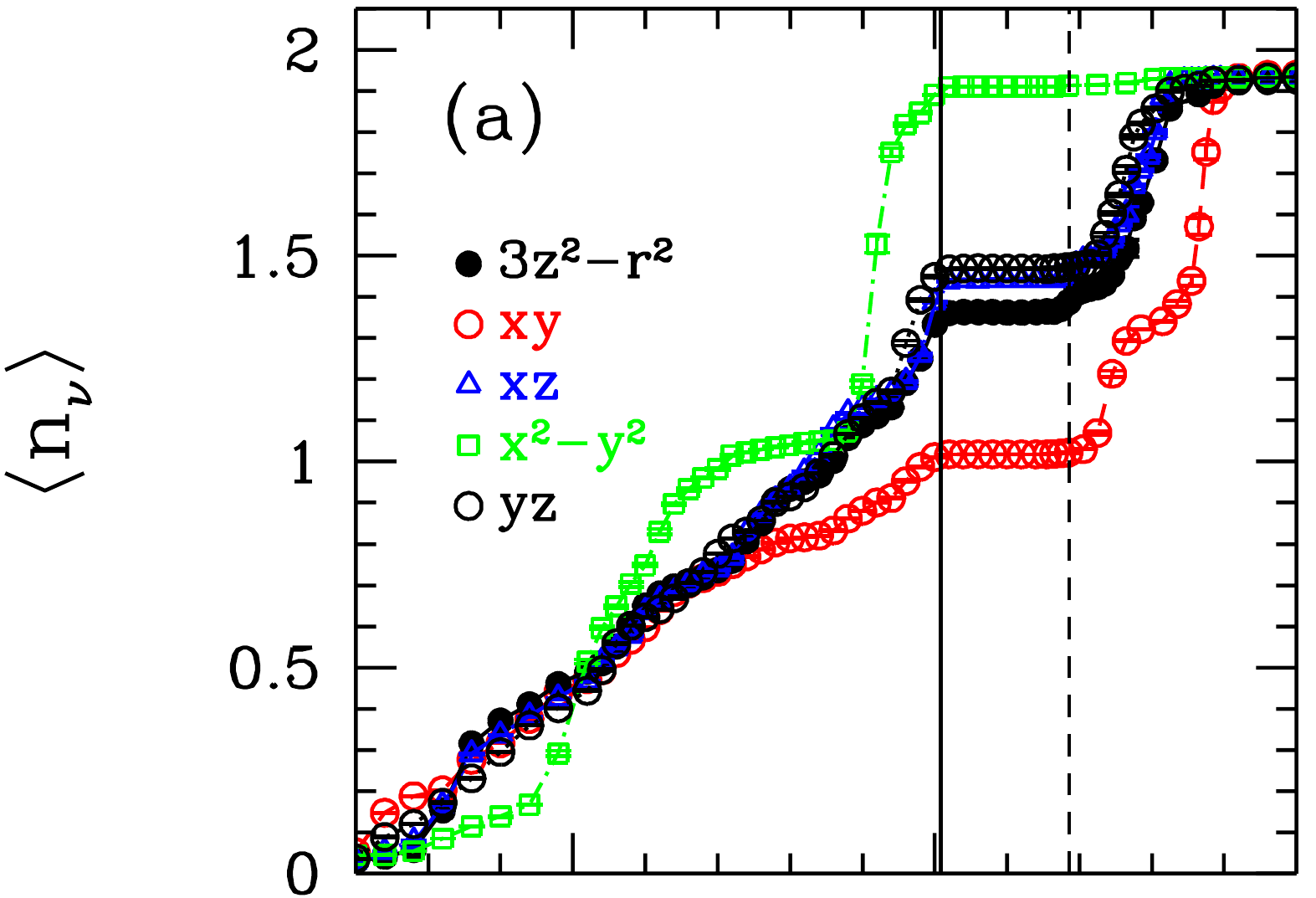} 
\includegraphics[width=6.8cm]{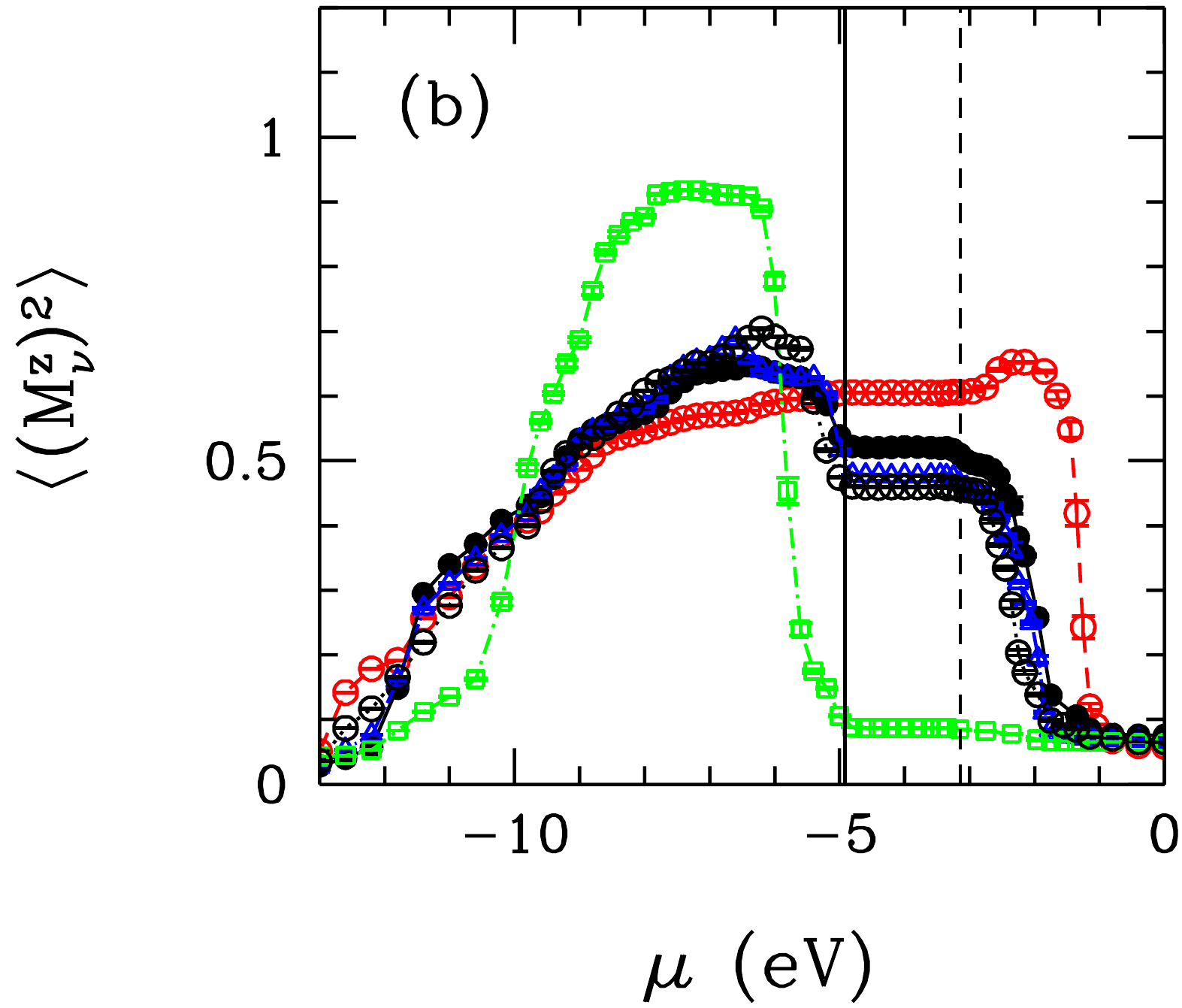} 
\caption{(Color online) 
(a) Electron occupation number $ \langle n_{\nu} \rangle $ 
of the Co($3d_{\nu}$) natural atomic orbitals 
versus the chemical potential $ \mu $.
(b) Square of magnetic moment $ \langle \left( M_{\nu}^{z}\right)^{2} \rangle $ 
for Co($3d_{\nu}$) natural atomic orbitals versus the chemical potential $ \mu $. 
Here, the vertical solid and dashed lines denote the values of the HOMO and LUMO levels, 
respectively. These results are for $ U=4 $ eV.
}
\label{fig5}
\end{figure}

Figure \ref{fig5}(a) shows 
the electron occupation number $\langle n_{\nu}\rangle = \sum_{\sigma} \langle d^{\dagger}_{\nu \sigma} d_{\nu \sigma} \rangle $ for the Co($3d_{\nu}$) NAO 
states as a function of the chemical potential $\mu$. 
In Fig. \ref{fig5}(a), we see that 
$\langle n_{\nu}\rangle$ becomes finite  
at $\mu \approx -12$ eV, and the Co($3d_{\nu}$) NAO's become singly occupied at $\mu \approx -6.5$ eV. 
At $\mu \approx -5.5$ eV, the Co($3d_{x^2-y^2}$) NAO becomes doubly occupied. At the HOMO level, 
the $3d_{xy}$ NAO is singly occupied, while the remaining orbitals have occupation number 
near $1.4$. Between the HOMO and LUMO levels, the Co($3d_{\nu}$) occupations do not change. 
The occupation of the Co($3d_{xy}$) NAO exhibits a sudden increase at $\mu \approx -2.5$ eV by 
about 0.3 electrons. We think that this sudden increase corresponds to an impurity bound state 
located at this energy. When $\mu$ reaches $-1.0$ eV, all of the Co($3d_{\nu}$) orbitals become doubly 
occupied. 

An interesting observation in Fig. \ref{fig5}(a) is that the impurity bound states are not induced in the semiconducting 
energy gap between the HOMO and LUMO levels. They are located above the LUMO level between $-3.0$ eV and $-2.0$ eV. Now, as seen in Fig. 
\ref{fig2}, the main features of the host density of states $D_{h}(\varepsilon)$ consist of a continuous 
conduction band located below $\approx -5 $ eV, two discrete states located at $-3.3$ eV and $-2.6$ eV, 
and a continuum of valence band states above $-2.0$ eV. In Fig. \ref{fig5}(a), we observe that 
the new impurity bound states are induced right below the continuum of valence band states located 
above $-2.0$ eV. The impurity bound states are located above the LUMO level because the 
spectrum of the host states is discrete around the LUMO level. 

Figure \ref{fig5}(b) shows the square of the magnetic moment at the Co($3d_{\nu}$) NAO's, $\langle (M^z_{\nu})^2\rangle$,  
where $M^z_{\nu} =  d^{\dagger}_{\nu \uparrow} d_{\nu \uparrow} - d^{\dagger}_{\nu \downarrow} d_{\nu \downarrow}  $, 
as a function of $\mu$. We observe that the magnetic moments of the Co($3d_{\nu}$) NAO's increase continuously, as $\mu$ is increased up to 
$\mu \approx -6.5$ eV. The magnetic moment of the Co($3d_{x^2-y^2}$) NAO decreases rapidly at $ \mu \approx -5.5$ eV 
due to double occupancy. For $ \nu = 3z^2-r^2$, $xz$ and $yz$,
the magnetic moments decrease in the interval $-6.5$ eV $ \ltsim \mu \ltsim -5.0$ eV, 
however, they do not go to zero. They have finite magnetic moments at the HOMO level. Between 
the HOMO and LUMO levels, the magnetic moments do not change. For the Co($3d_{xy}$) NAO, 
$\langle (M^{z}_{\nu})^2 \rangle$ exhibits a small increase at $\mu \approx -2.5$ eV. Upon 
further increase of $\mu$, for all of the Co($3d_{\nu}$) NAO's, $\langle (M^{z}_{\nu})^2 \rangle $ vanishes 
as the orbitals become doubly occupied. 

\begin{figure}
\includegraphics[width=6.8cm]{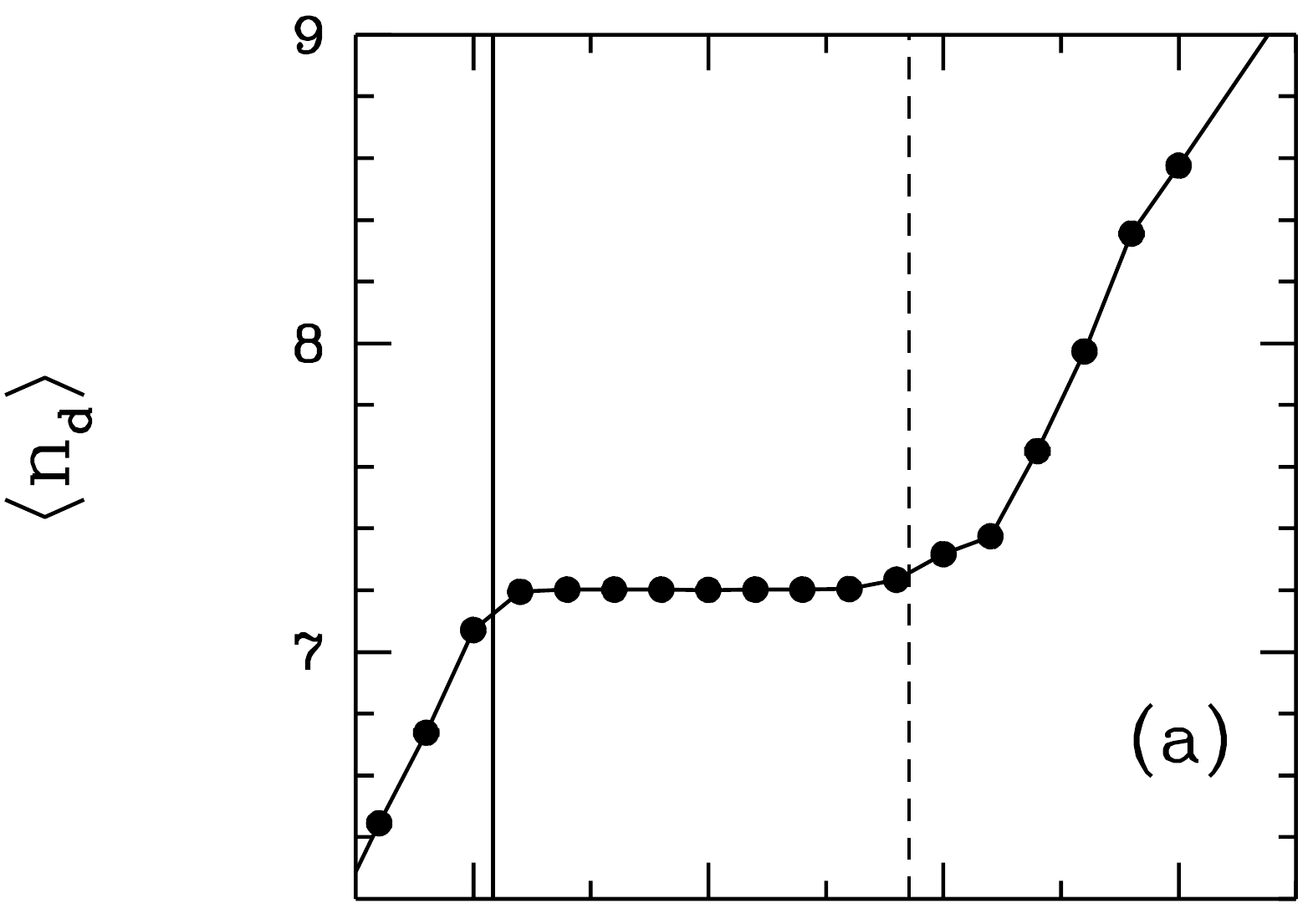}
\includegraphics[width=6.8cm]{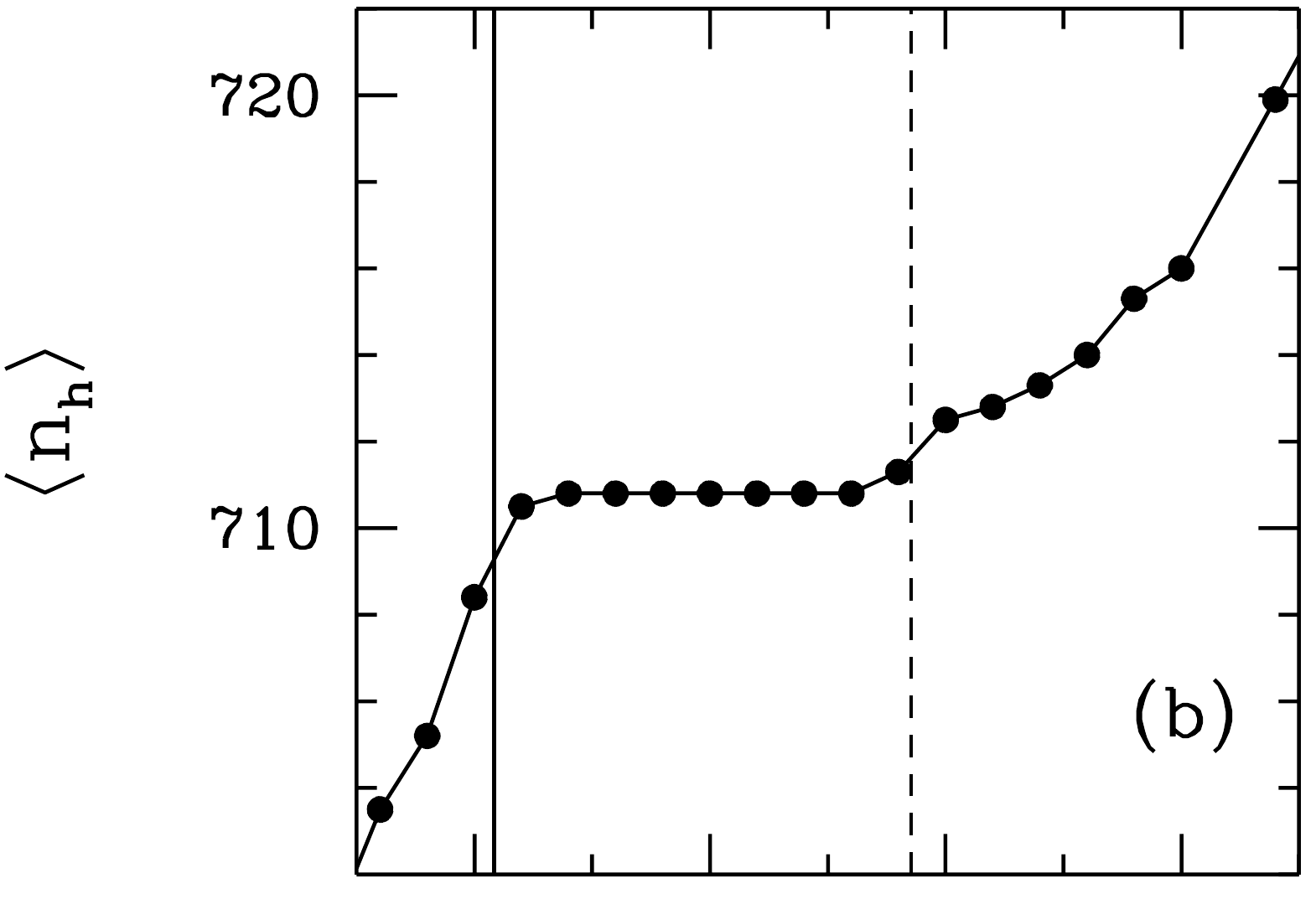}
\includegraphics[width=6.8cm]{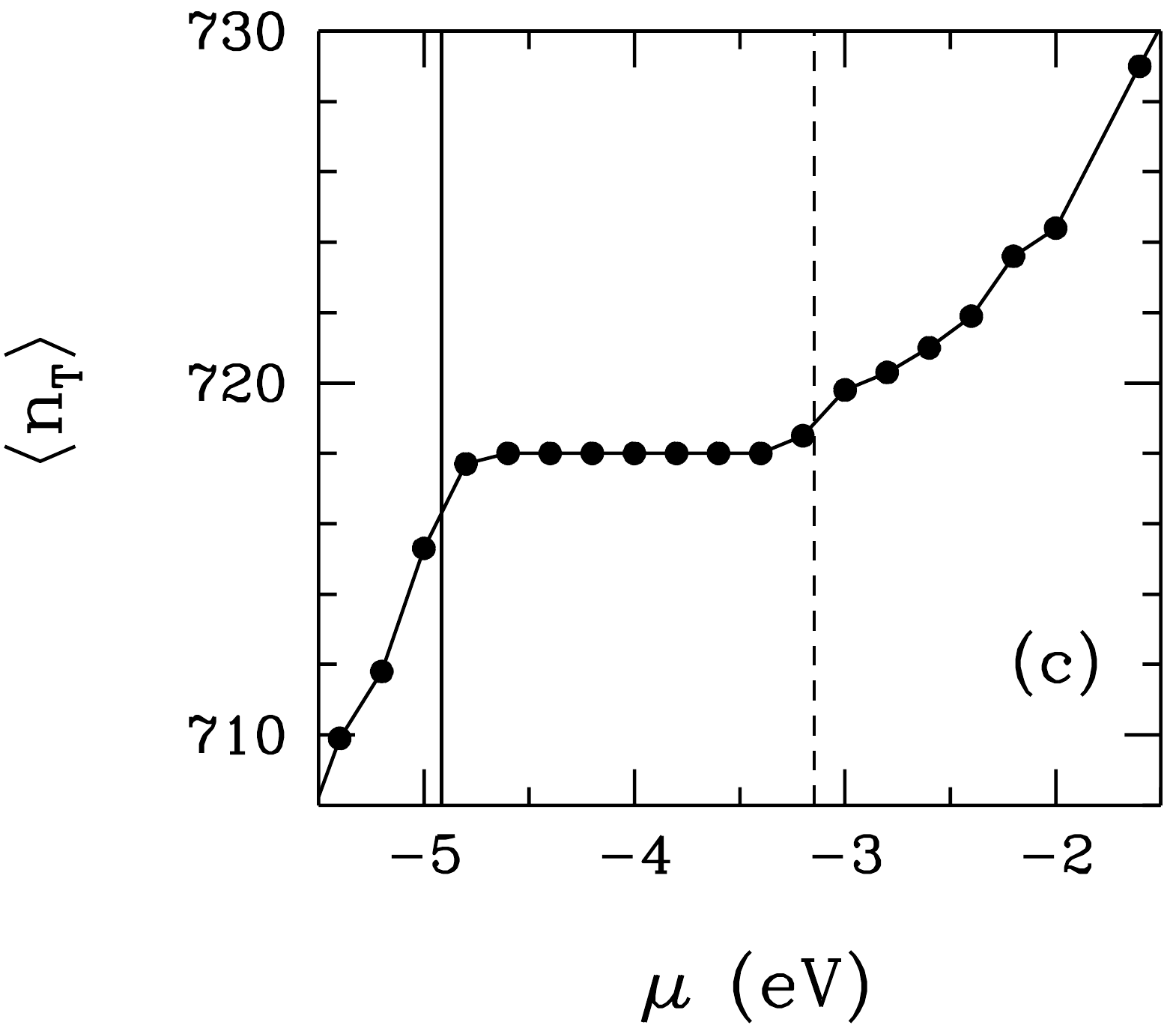}
\caption{(Color online)  
(a) Total electron occupation number $\langle n_{\mathrm{d}} \rangle$ of the Co($3d$) natural atomic orbitals versus chemical potential $\mu$. 
(b) Total number of the host electrons $\langle n_{\mathrm{h}} \rangle$ versus $\mu$.
(c) Total number of electrons $\langle n_{\mathrm{T}} \rangle= \langle n_{\mathrm{d}} \rangle + \langle n_{\mathrm{h}} \rangle$ for CNCbl versus $\mu$.  
Here, 
the vertical solid and dashed lines denote the HOMO and LUMO levels, 
respectively. The charge neutral CNCbl molecule contains 718 electrons.
These results are for $U=4$ eV.
}
\label{fig6}
\end{figure}

In Fig. \ref{fig6}(a), we present QMC data on the total electron occupation of the Co($3d_{\nu}$) NAO's 
$\langle n_{\mathrm{d}} \rangle = \sum^{5}_{\nu=1} \sum_{\sigma} \langle d_{\nu \sigma}^{\dagger} d_{\nu \sigma} \rangle$ 
as a function of $\mu$. We see that $\langle n_{\mathrm{d}} \rangle$ increases up to HOMO level, it equals 7.2 at 
$\mu \approx -4.8$ eV.  We observe that $\langle n_{\mathrm{d}} \rangle$ does not change between the HOMO and 
LUMO levels.  
The total number of the host electrons
$\langle n_{\mathrm{h}} \rangle =  \sum^{N-5}_{m=1} \sum_{\sigma} \langle c_{m \sigma}^{\dagger} c_{m \sigma} \rangle$ is 
shown in Fig. \ref{fig6}(b). 
Figure \ref{fig6}(c) shows the total electron number for CNCbl $\langle n_{\mathrm{T}} \rangle = \langle n_{\mathrm{d}} \rangle + \langle n_{\mathrm{h}} \rangle$   
versus $\mu$. Here, we clearly see that between the HOMO and LUMO levels the total electron number $\langle n_{\mathrm{T}} \rangle = 718$ 
corresponding to the neutral CNCbl molecule.
\begin{figure}
\includegraphics[width=6.8cm]{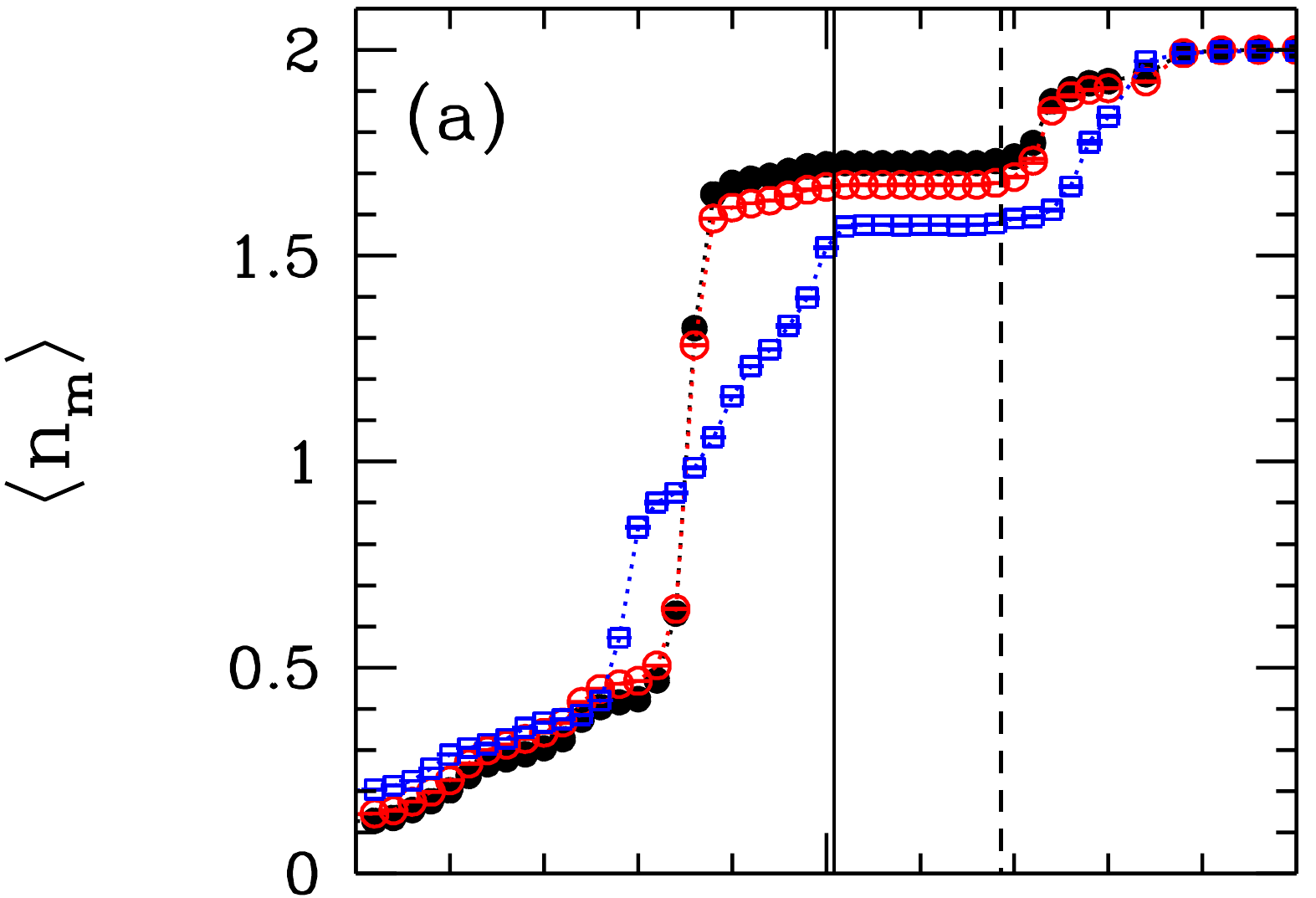}
\includegraphics[width=6.8cm]{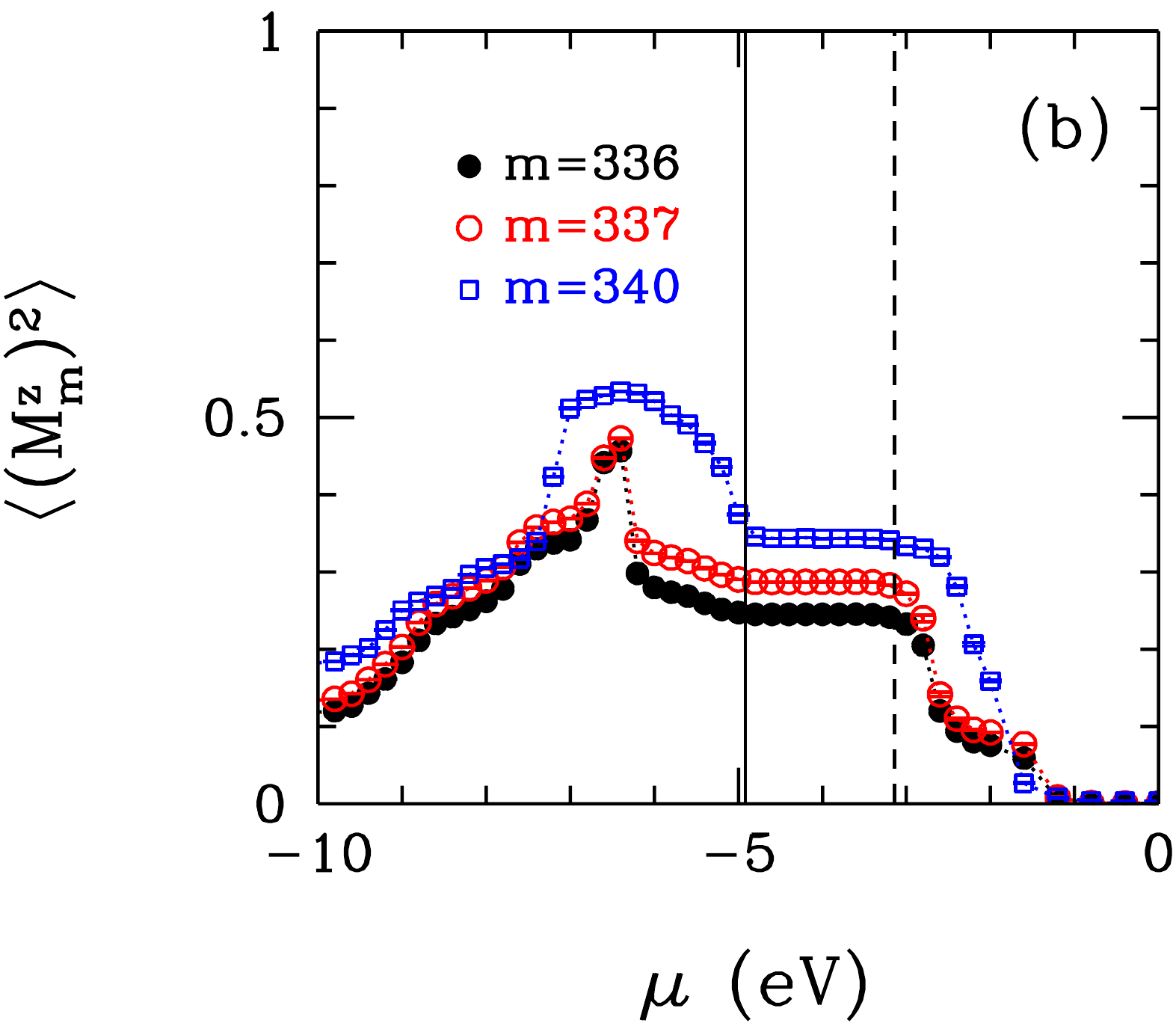} 
\caption{(Color online)  
(a) Electron occupation number of the $m$'th host state $\langle n_m\rangle$ versus $\mu$.
(b) Square of the magnetic moment of the $m$'th host state 
$\langle (M_m^z)^2 \rangle$ versus $\mu$.
Here, 
the vertical solid and dashed lines denote the HOMO and LUMO levels, 
respectively.
These results are for $U=4$ eV.
}
\label{fig7}
\end{figure}

In Fig. \ref{fig7}(a), we present QMC data on the host electron number $\langle n_{m} \rangle = \sum_{\sigma} \langle c^{\dagger}_{m \sigma} c_{m \sigma}
\rangle$ versus $\mu$ for the $m=336, 337$ and $340$ host eigenstates. The bare energy levels $\varepsilon_{m}$ of 
these states are located at $-6.48$ eV, $-6.43$ eV and $-6.16$ eV for  $m=336, 337$ and $340$, respectively. 
Here, we observe that these host states do not become doubly occupied as $\mu$ passes through 
the $\varepsilon_{m}$'s. For example, at the HOMO level, $\langle n_{m} \rangle = 1.72$, $1.67$ and 
$1.52$ for $m=336, 337$ and $340$ respectively even though they are located deep below the HOMO level. Consequently, these host states have finite 
magnetic moments when $\mu$ is at the HOMO level as seen in Fig. \ref{fig7}(b). The magnetic moments vanish 
after these host states become doubly occupied for $\mu \gtsim -1$ eV. 

\begin{figure}[!htpb]
\includegraphics[width=6.8cm]{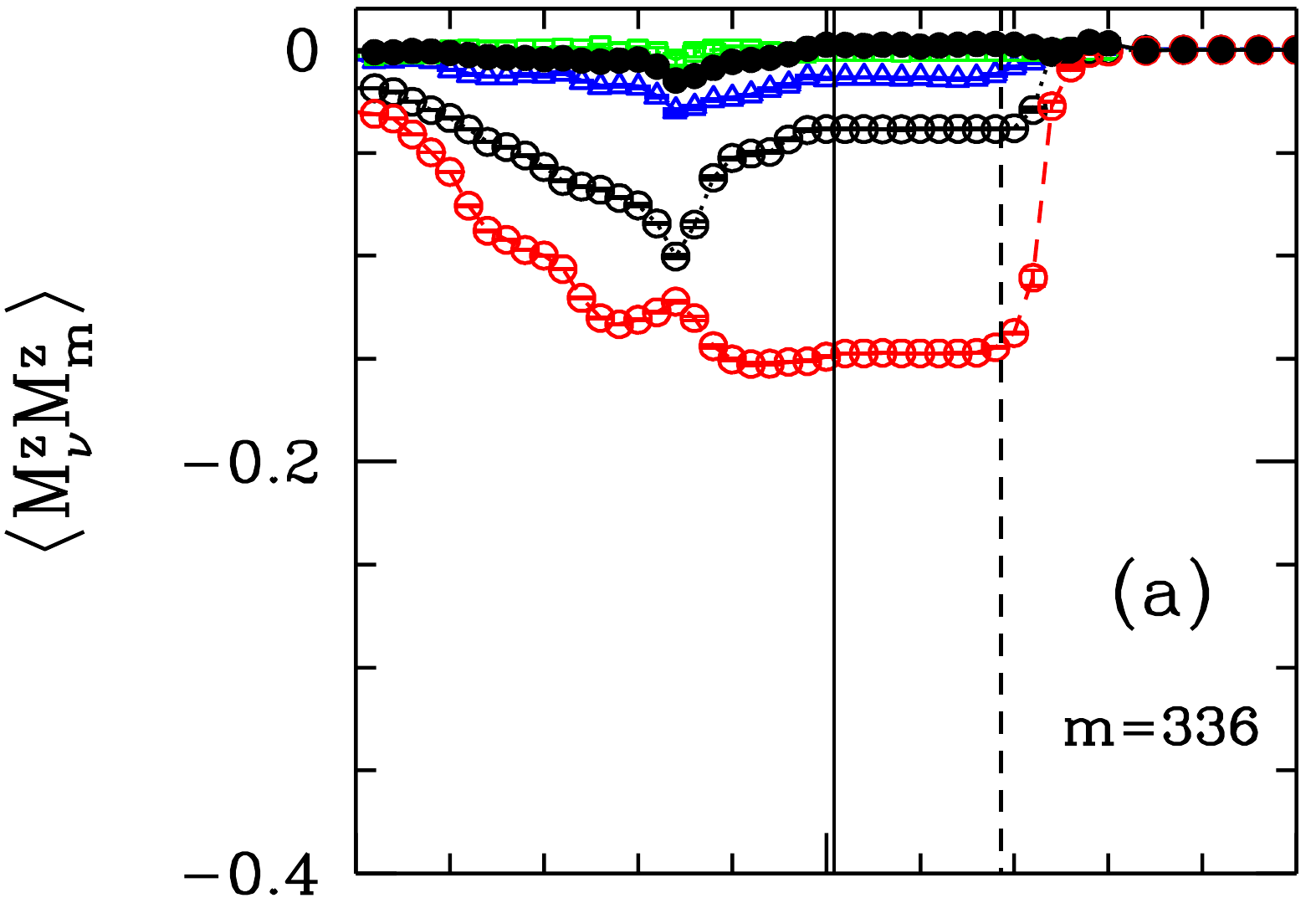}
\includegraphics[width=6.8cm]{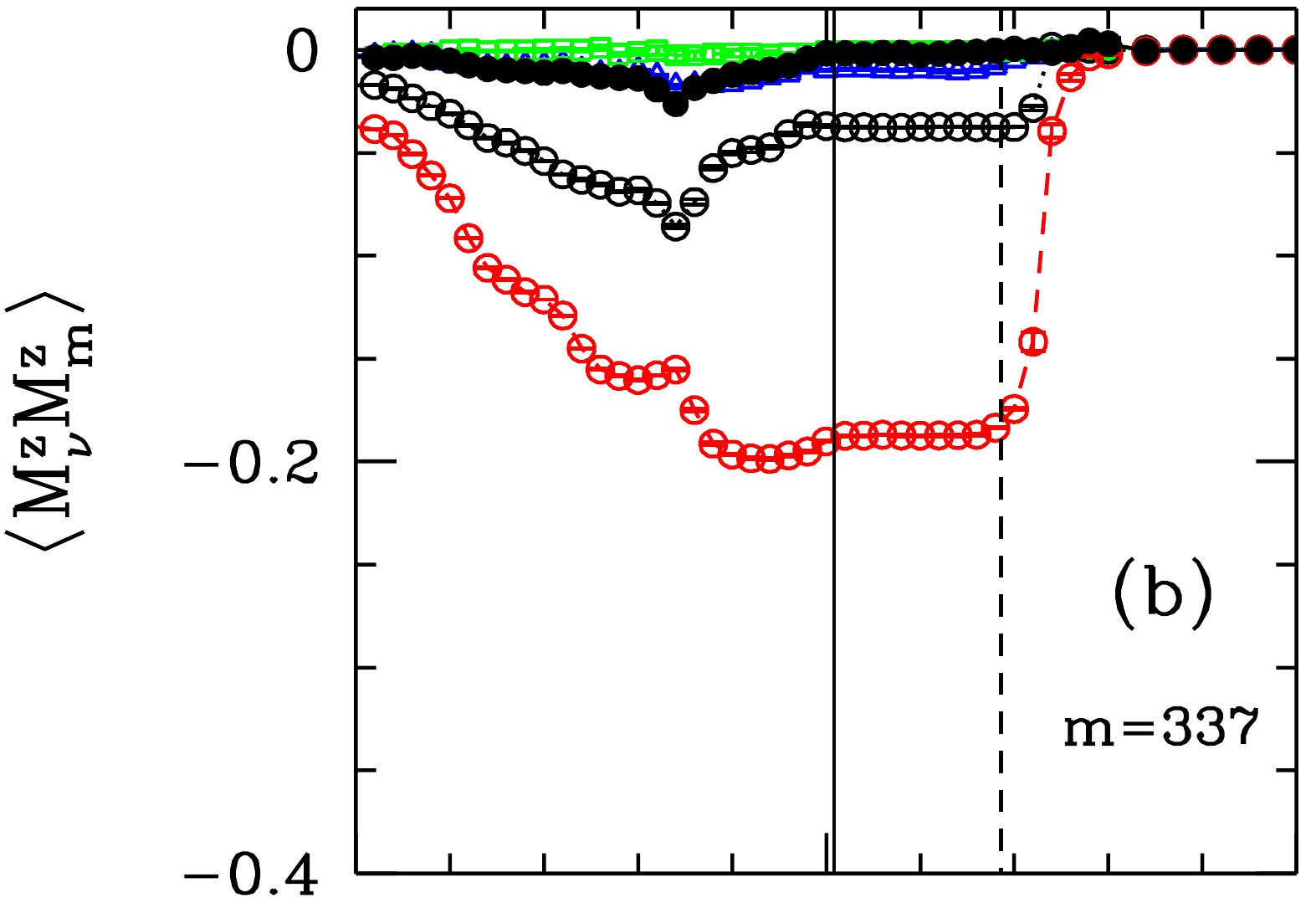}  
\includegraphics[width=6.8cm]{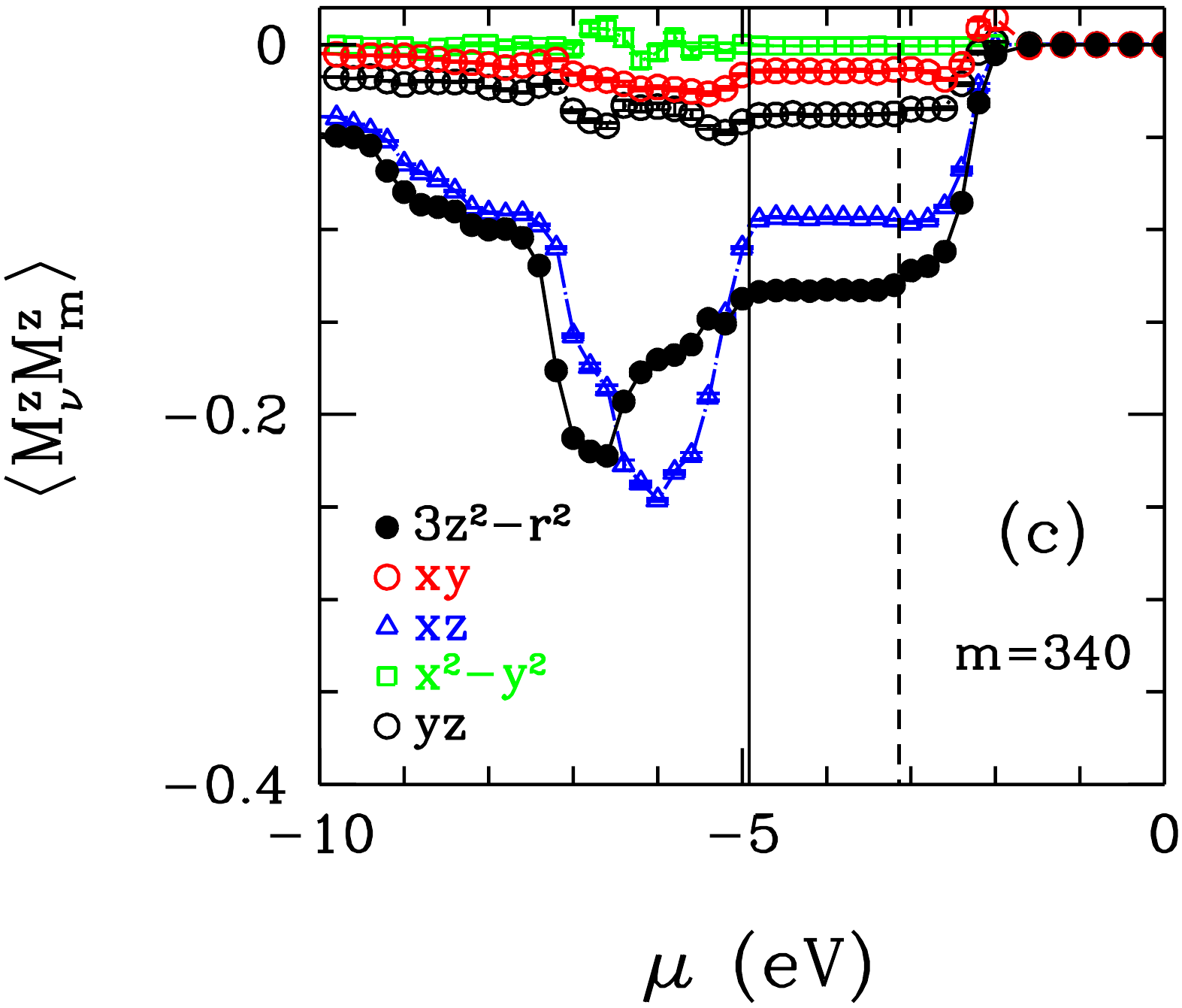}
\caption{(Color online)  
Magnetic correlation function
$\langle M_{\nu}^z M_m^z \rangle$
between the $m$'th host state and 
the Co($3d_{\nu}$) natural atomic orbitals.
Here, results are shown for the host states 
(a) $m=336$, (b) 337, and (c) 340. 
The vertical solid and dashed lines denote the HOMO and LUMO levels, 
respectively.
These results are for $U=4$ eV.
}
\label{fig8}
\end{figure}

Next, in Fig. \ref{fig8}(a)-(c), we present QMC data on the magnetic correlation function 
$\langle M^{z}_{\nu} M^{z}_{m} \rangle$ between the magnetic moments at the Co($3d_{\nu}$) NAO's and the 
$m$'th host states for $m=336, 337$ and $340$. These figures show that the host states with the strongest 
hybridization have antiferromagnetic correlations with the moments at the Co($3d_{\nu}$) NAO's. 
These antiferromagnetic correlations vanish as the host states become doubly occupied. 
 
\section{Comparison of the HF+QMC and the DFT+QMC results}
In Ref. [\onlinecite{Kandemir2}], numerical results were presented from previous HF+QMC calculations 
on Im-[Co$^{\rm III}$(corrin)]-CN$^+$, which is a smaller piece of the CNCbl molecule. Here, we 
compare those results with the current DFT+QMC data obtained for the whole CNCbl molecule. 

We find various differences between the outcomes of the HF and DFT calculations. In the HF method, 
non-interacting electrons are described under the influence of a mean field potential 
which consists of the classical Coulomb potential and a non-local exchange potential. On the 
other hand, in the DFT calculations a local exchange potential is used \cite{Harrison}. Furthermore, 
in the HF+QMC approach, the intra-orbital Coulomb interaction was assumed to be unscreened and the 
bare Coulomb matrix elements were used. For cobalt, the intra-orbital $U$ was taken to be about $36$ eV. 
On the other hand, in the DFT+QMC approach, the intra-orbital Coulomb interaction is assumed to be renormalized due to long-range 
screening effects. Here, $U$ was taken to be $4$ eV. 

In the HF calculations, the locations of the Co($3d_{\nu}$) NAO's separate into two groups 
corresponding to the $e_{g}$ and $t_{2g}$ symmetries. In the HF+QMC calculations, we have used a constant 
energy shift $\mu^{\mathrm{DC}}$ to compensate for the double counting of $U$. The resultant QMC data reflect 
this ordering of the Co($3d_{\nu}$) NAO's. However, the DFT calculations yield nearly degenerate Co($3d_{\nu}$) 
NAO's. In the DFT+QMC approach, we have used an orbital dependent $\mu_{\nu}^{\mathrm{DC}}$ which led 
to the ordering of the Co($3d_{\nu}$) NAO's seen in Fig. \ref{fig2}(b). We find that there are some differences 
in the locations of the shifted Co($3d_{\nu}$) NAO's where we compare the HF and DFT results. These differences clearly influence the outcome of the 
QMC calculations. Nevertheless, we observe that impurity bound states involving the Co($3d_{xy}$) and 
Co($3d_{3z^{2}-r^{2}}$) NAO's are found in both the HF+QMC and DFT+QMC calculations. In both cases, the impurity 
bound state for the Co($3d_{3z^{2}-r^{2}}$) NAO is located higher in energy compared to that of the Co($3d_{xy}$). 

\section{Summary and conclusions}
In summary, we have studied the electronic structure and magnetic correlations of cyanocobalamin. For this purpose, we have used the 
multi-orbital single-impurity Halda\-ne-\-An\-der\-son model of a transition metal impurity embedded in a semiconductor host. 
First, we have constructed an effective Haldane-Anderson model by using DFT calculations.
We have obtained the one-electron  
parameters of this model from the Kohn-Sham matrix written in the basis of the natural atomic orbitals. 
We have taken $U=4$ eV and have shifted the Co$(3d_{\nu})$ levels by $\mu^{\mathrm{DC}}_{\nu}$ 
to prevent the double-counting of the intra-orbital Coulomb interaction by both DFT and QMC. 

The QMC results clearly show how the single-electron spectral weight is distributed in energy. 
In our calculations, we see that as the chemical potential increases, the Co($3d_{\nu}$) NAO's become occupied. When the chemical potential 
re\-ac\-hes the HOMO level, the total electron number equals 718. This corresponds to the neutral CNCbl molecule. 
In this case, the Co($3d_{\nu}$) NAO's are less than doubly occupied and they have finite magnetic moments. 
Between the HOMO and LUMO levels, there is no single-particle spectral weight. We observe that 
above the LUMO level and between $-3.0$ eV and $-2.0$ eV, there are new states induced by the Coulomb 
interaction $U$. We identify these new states as impurity bound states because of the filling 
dependence of the antiferromagnetic correlations between the Co($3d_{\nu}$) NAO's and the host magnetic 
moments. This identification is similar to that done previously in the analysis of the HF+QMC 
results presented in Ref. [\onlinecite{Kandemir2}]. 
The impurity bound state is most clearly seen for the Co($3d_{xy}$) NAO. Surprisingly according to the DFT+QMC result, the impurity bound 
states are located above the LUMO level instead of being in the semiconducting energy gap. We think that this is 
because of the discrete energy spectrum of the CNCbl molecule. 

It remains to be seen whether the impurity 
bound states found in the DFT+QMC calculations are related to the peaks observed in the photoabsorption 
spectrum of CNCbl. For a more direct comparison with the experimental data, it would be necessary to include the inter-orbital Coulomb 
interactions along with the Hund's coupling. We note that we have perfomed similar DFT+QMC calculations for 
hemoglobin, where we also find impurity bound states. Hence, these correlated electronic states appear to be a common feature of 
metalloproteins and metalloenzymes. We think that it will be interesting to figure out whether the impurity 
bound states have a general role in the functioning of metalloproteins and metalloenzymes. 

\begin{acknowledgements}

We thank
Hadi M. Zareie, Tahir \c{C}a\u{g}{\i}n, Mehmet Sar{\i}kaya, 
Nuran Elmac{\i}, \"{O}zg\"{u}r \c{C}ak{\i}r,
Devrim G\"{u}\c{c}l\"{u}, 
Jingyu Gan, Bo Gu, and Sadamichi Maekawa 
for valuable discussions and suggestions. 
The numerical calculations reported here were performed 
in part at the TUBITAK ULAKBIM, High Performance and Grid
Computing Center (TRUBA resources).
Financial support by the Turkish Scientific and Technical Research Council 
(TUBITAK grant numbers 110T387 and 113F242) is gratefully acknowledged. 

\end{acknowledgements}

\bibliographystyle{apsrev4-1}


\begin{thebibliography}{999}
 \bibitem{Prada} Pratt, J.M., Inorganic Chemistry of Vitamin B$_{12}$ (Academic Press,1972).

\bibitem{Harris} Harris, D.Ahmasi, Stickrath, Andrew B., Carroll, Elizabeth C., Sension, Roseanne J.: Influence of environment on the electronic structure of
Cob(III)alamins: Time-resolved absorption studies of the S$_{1}$ state spectrum and dynamics. J. Am. Chem. Soc. {\bf 129}, 7578-7585 (2007).

\bibitem{Stich} Stich, Troy A., Brooks, Amanda J., Buan, Nicole R., Brunold, Thomas C.: Spectroscopic and computational studies of Co$^{+3}$-Corrinoids:
Spectral and electronic properties of the B$_{12}$ cofactors and biologically relevant precursors. J. Am. Chem. Soc. {\bf 125}, 5897-5914 (2003).

\bibitem{Firth}  Firth, R. A., Hill, H. A. O. , Pratt, J. M. ,Williams, R. J. P. ,Jackson, W. R.: The circular dichroism and absorption spectra of some Vitamin B$_{12}$ derivatives.
Biochemistry {\bf 6}, 2178-2189 (1967). 

\bibitem{Grun} Grun, F., Menasse, R.: Estimation of the magnetic susceptibility of Vitamin B$_{12}$. Experientia {\bf 6}, 263-264 (1950).

\bibitem{Diehl} Diehl, H., Haar, R.W.V. and Sealock, R.R.: The magnetic susceptibility of Vitamin B$_{12}$. J. Am. Chem. Soc. {\bf 72},
5312-5313 (1950).

\bibitem{Haldane} Haldane, F.D.M., Anderson, P.W.: Simple model of multiple charge states of transition-metal impurities in semiconductors. Phys.Rev.B {\bf 13}, 2553 (1976).

\bibitem{Anderson} Anderson, P.W.: Localized magnetic states in metals.
Phys. Rev. {\bf 124}, 41 (1961).

\bibitem{Gaussian} Frisch, M. J., Trucks, G.W., Schlegel, H. B., Scuseria, G. E., Robb, M. A., Cheeseman, J.R., Scalmani, G., Barone, V., Mennucci, B., Petersson, G. A., Nakatsuji, H., Caricato, M., Li, X., Hratchian, H.P., Izmaylov, A.F., Bloino, J., Zheng, G., Sonnenberg, J.L., Hada, M., Ehara, M., Toyota, K., Fukuda, R., Hasegawa, J., Ishida, M., Nakajima, T., Honda, Y., Kitao, O., Nakai, H., Vreven, T., Montgomery, J.A.,Peralta, J. E., Ogliaro, F., Bearpark, M., Heyd, J.J., Brothers, E., Kudin, K.N., Staroverov, V.N., Kobayashi, R., Normand, J., Raghavachari, K., Rendell, A., Burant, J.C., Iyengar, S.S., Tomasi, J., Cossi, M., Rega, N., Millam, J.M., Klene, M., Knox, J. E., Cross, J.B., Bakken, V., Adamo, C., Jaramillo, J., Gomperts, R., Stratmann, R.E., Yazyev, O., Austin, A.J., Cammi, R., Pomelli, C., Ochterski, J.W., Martin, R.L., Morokuma, K., Zakrzewski, V.G., Voth, G.A., Salvador, P., Dannenberg, J.J., Dapprich, S., Daniels, A.D., Farkas, O., Foresman, J.B., Ortiz, J.V., Cioslowski, J., and  Fox, D.
J.:
Gaussian 09, Revision D.01,  Gaussian, Inc., Wallingford CT (2009).

\bibitem{Hirsch} Hirsch, J. E. and Fye, R. M.: Monte Carlo method for magnetic impurities in metals. 
Phys. Rev. Lett. {\bf 56}, 2521 (1986).

\bibitem{Kandemir2} Kandemir, Z., Mayda, S. and Bulut, N.: Electronic structure and correlations of Vitamin B$_{12}$ studied within the Haldane-Anderson impurity model.
European Physical Journal B {\bf 89}, 113 (2016).

\bibitem{Kohn} Kohn, W. and Sham., L.J.: Self-consistent equations including exchange and correlation effects. Phys. Rev. {\bf 140}, A1133 (1965).

\bibitem{Reed} Reed, A.E., Curtiss, L.A. and Weinhold, F.: Intermolecular interactions from a natural bond orbital, donor-acceptor viewpoint. Chem. Rev. {\bf 88}, 899-926 (1988).

\bibitem{BP86} Becke, A.D.: Density-functional exchange-energy approximation with correct asymptotic behavior.
Phys. Rev. A {\bf 38}, 3098 (1988); 
Perdew, J.P.: Density-functional approximation for the correlation energy of the inhomogeneous electron gas.
Phys. Rev. B {\bf 33}, 8822 (1986).

\bibitem{Sasioglu} \c{S}a\c{s}{\i}o\u{g}lu, E., Galanakis, I., Friedrich, C. and Bl\"{u}gel, S.: Ab initio calculation of the effective on-site Coulomb interaction parameters for half-metallic magnets. Phys. Rev. B {\bf 88}, 134402 (2013).

\bibitem{Anisimov} Anisimov, V.I., Zaanen, J., Andersen, O.K.: Band theory and Mott insulators: Hubbard U instead of Stoner I.
Phys. Rev. B {\bf 44}, 943 (1991).

\bibitem{Czyzyk} Czyzyk, M.T. and Sawatzky, G.A.: Local-density functional and on-site correlations: The electronic structure of La$_{2}$CuO$_{4}$ and LaCuO$_{3}$.
Phys. Rev. B {\bf 49}, 14211 (1994).

\bibitem{Kunes} Kune{\~s}, J., Anisimov, V.I., Lukoyanov, A.V. and Vollhardt, D.: Local correlations and hole doping in NiO: A dynamical mean-field study.
Phys. Rev. B {\bf 75}, 165115 (2007). 

\bibitem{Karolak} Karolak, M., Ulm, G., Wehling, T. O., Mazurenko, V.,Poteryaev, A. and Lichtenstein, A.: Double counting in LDA + DMFT$-$The example of NiO. 
Journal of Electron Spectroscopy and Related Phenomena  
{\bf 181}, 11-15 (2010).

\bibitem{Harrison} Harrison, N. M.: An introduction to density functional theory. Nato Science Series Sub Series III Computer and Systems Science {\bf 187}, 45-70 (2003).


\end{thebibliography}

\end{document}